\documentclass[10pt]{article}
\usepackage{graphicx}
\usepackage{caption}
\usepackage{subcaption}
\usepackage{amsfonts}
\usepackage{amsmath}
\usepackage{amssymb}
\usepackage[margin=2.00cm]{geometry}

\begin{document}

\title{Direct and indirect seismic inversion: interpretation of certain mathematical theorems}

\author{August Lau and Chuan Yin}

\maketitle

\begin{abstract}
Quantitative methods are more familiar to most geophysicists with direct inversion or indirect inversion.  We will discuss seismic inversion in a high level sense without getting into the actual algorithms.   We will stay with meta-equations and argue pros and cons based on certain mathematical theorems.
\end{abstract}

\section*{Introduction}

Seismic inversion can be divided into QUANTITATIVE and QUALITATIVE methods.  We will only discuss quantitative methods in this paper.  And we will not address specific implementations.   So the arguments would be more at the operator level rather than numerical level.

\section*{Quantitative method}

We will first discuss QUANTITATIVE METHOD.  Seismic inversion (inverse problem) can be roughly divided into direct method and indirect method.  The general meta-equation would be
$$ Ax = y, $$
where $A$ is the operator, 
$x$ is the solution, 
$y$ is the input data.

Direct inversion is to find an operator B so that
$$ x = By, $$
where $A$ and $B$ are roughly inverse to each other.

Indirect inversion is to find $x$ so that certain optimization criterion is satisfied.
Typically it would be minimization of $L2$ $(Ax-y)$ where $L2$ is the least square norm.

\section*{Direct inversion (GAUSS)}

Direct inversion is probably the ''oldest'' method of inversion.  Let us take a 1-D example of convolution,
\begin{equation} \label{conv}
 W*R = D, 
\end{equation}
where $W$ is the wavelet,  $R$ is reflectivity and $D$ is data trace.
Let us assume that $W$ is known and the inversion seeks $R$ the unknown.
$$ W(z)R(z) = D(z) $$
if we use the $z$-transform approach.
By the Fundamental Theorem of Algebra (attributed to GAUSS),
$$ W(z) = a(z-a_1)(z-a_2)...(z-a_n), $$
but it is an existence theorem.

\subsection*{Reality check (real data)} 

For real data like field recorded data,  there are attendant questions of how many terms, how accurate is each monomial inverse, ''noise'' and artifacts, etc.  So it is not as seamless as factorization or series expansion.

\section*{Galois theory}

Galois group theory defines the concept of what is solvable by radicals.  It means that we could only use square, square root, cube, cube root, etc, of the original coefficients.   Not every polynomial can be factorized in simple monomials by using only radicals.  In fact, if the polynomial is order 5 or above,  there will be cases that they could not be solvable (by radicals).   

For quadratic equation (second power), we can decompose it into simple monomials with quadratic equation.  For quintic and above (fifth power),  there is no general formula.  Then the inverse operator would have to be approximated.  This is a theoretical hint from Galois Theoerem that solvability of the exact solution cannot be found.

\section*{Indirect inversion (attributed to GAUSS)}

Indirect inversion uses the forward modeling to fit data.  The most common method is least square (attributed to GAUSS), by minimizing 
$$ \|W*R - D\|_{L_{2}}. $$

\subsection*{Reality check (real data)}

For real data like field recorded data, there are attendant questions of convergence, regularization, prior information like well control, low and high frequency instability, etc.
So it is not clear how to do the optimization which could range from gradient to random search methods. It is still a trial-and-error approach to select algorithms.

\section*{Further discussions}

\subsection*{Indirect inversion advantages}

Indirect inversion is able to include prior information like well logs, gravity, magnetic, preliminary interpretation, geometric measures, etc. These other data types could be included in the optimization so the solution could fit the seismic as well as the prior information.

Indirect inversion could use L0, L1 and other norms which help stability by enforcing sparsity.  Also it is easy to incorporate ''geometric inversion'' which solves for boundary geometry like salt, carbonate etc.  Examples of geometric inversion are DLT (deformable layer tomography) (Zhou 2006) and level set method (Lewis and Vigh, 2016).

A good summary of quantitative method could be found in ''Direct and indirect inversion'' by Virieux et al. (2015);  also ''Inverse problems in classical and quantum physics'', by Almasy (2009), has made numerical studies of non-uniqueness of inverse problem.  In particular,  the Riemann-Lebesgue Lemma shows that the null-space could have a lot of solutions which yield same modeled data up to a small epsilon.  For real field data,  we could assume that the real data is the same as modeled data when they match up to an epsilon.

\subsection*{A Riemann-Lebesgue example}

The Riemann-Lebesgue Lemma states that if the Lebesgue integral of $f$ is finite, then the Fourier transform of $f$ satisfies
$$
F(k):=\int_{R^d} f(x) exp(-ikx)dx \Rightarrow 0, 
k \Rightarrow \infty.
$$
This is equivalent to say the following for 1-D real functions $f:[a,b] \rightarrow R$ 
$$ \lim_{k \rightarrow \infty} \int^b_a f(x) \sin (kx) dx = \lim_{k \rightarrow \infty} \int^b_a f(x) \cos (kx) dx = 0 $$

A practical implication of the Riemann-Lebesgue Lemma on geopysical inversion can be demonstrated with the familiar 1-D convolutional model, Equation (\ref{conv}), which can be written as the  following
$$ d(t) = w(t) * r(t) = \int^t_0 w(t-\tau) r(\tau) d\tau , $$ 
where $w(t)$ is the wavelet function, $r(t)$ the reflectivity function, and $d(t)$ the seismogram. Acoording to the Riemann-Lebesgue Lemma, a different reflectivity function would also yield the identical seismogram,
$$ r_1 (t) = r(t) + \alpha \sin (\beta t) ,$$
because 
\begin{eqnarray*}
\lim_{\beta \rightarrow \infty} \int^t_0 w(t-\tau) [ r(\tau) + \alpha \sin (\beta \tau) ] d\tau & = &  \\
                                                               \int^t_0 w(t-\tau) r(\tau) d\tau & = & d(t). 
\end{eqnarray*}
\begin{figure}
\centering
  \includegraphics[width=3in]{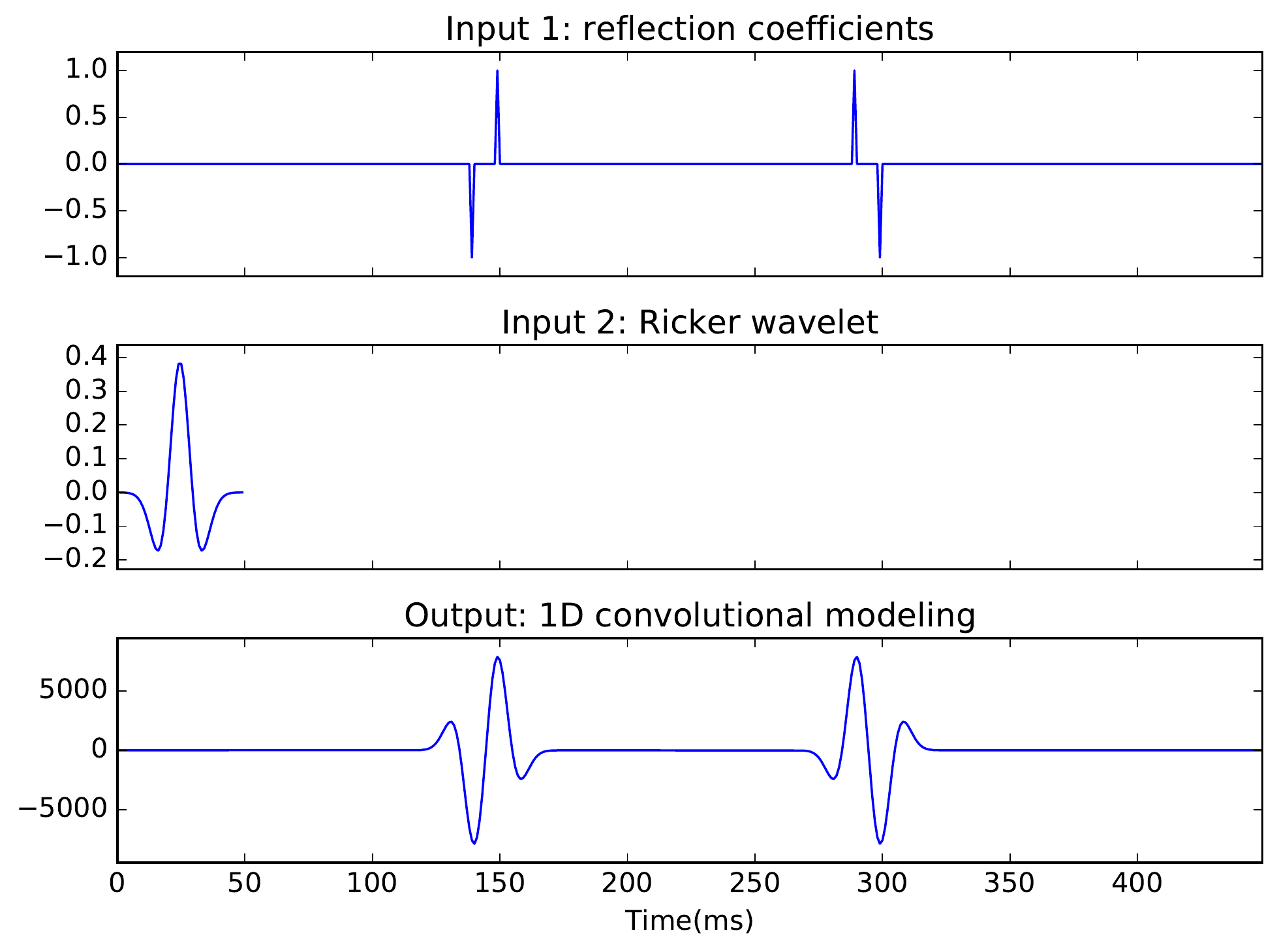}
\caption{Forward modeling using 1D convolutional modeling, $\alpha=0.0$, $\beta=0.0$}
\end{figure}
\begin{figure}
\centering
  \includegraphics[width=3in]{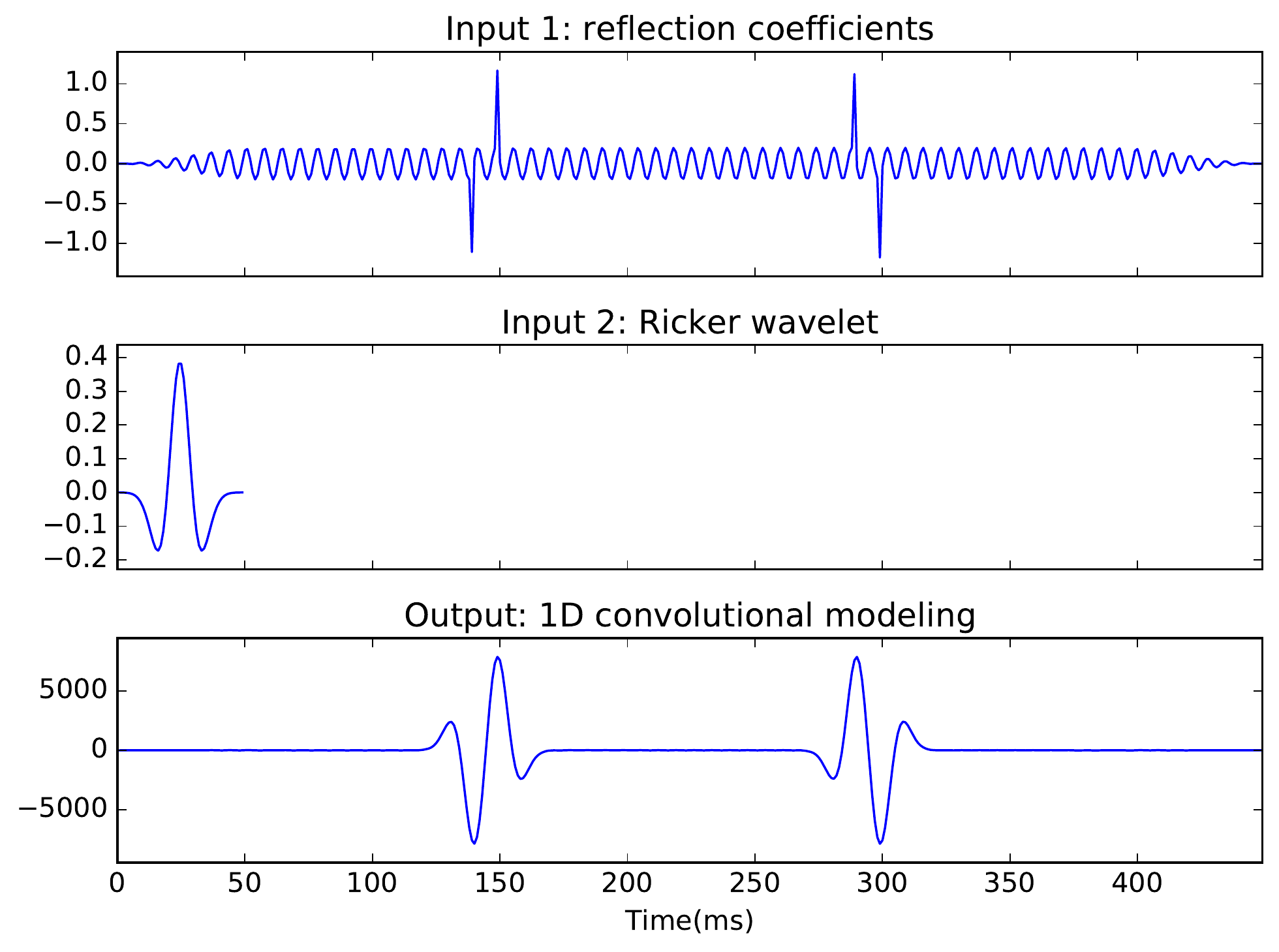}
\caption{Forward modeling using 1D convolutional modeling, $\alpha=0.2$, $\beta=0.9$}
\end{figure}
\begin{figure}
\centering
  \includegraphics[width=3in]{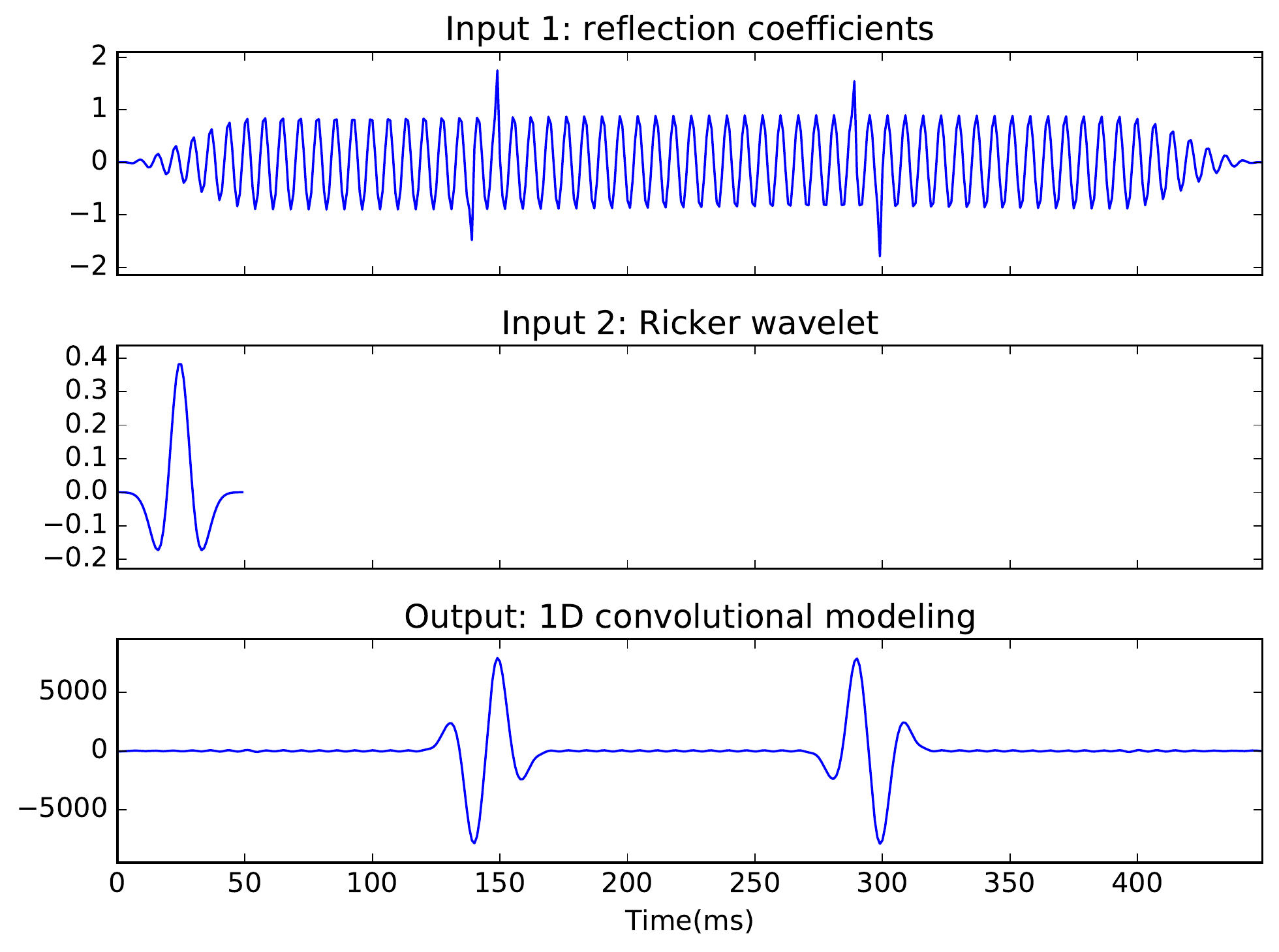}
\caption{Forward modeling using 1D convolutional modeling, $\alpha=0.9$, $\beta=0.9$}
\end{figure}

Figures 1 - 3 are numerical and graphical demonstrations of this analytical result, where Figure 1 uses the original reflectivity. Figures 2 and 3 use non-zero values for $\alpha$ and $\beta$, $\alpha=0.2$,$\beta=0.9$, and $\alpha=0.9$,$\beta=0.9$, respectively. This example is a reminder that there is a huge null space even for just a 1-D inversion using the convolutional model.

\subsection*{Some practical aspects}

Despite the significant efforts used in direct/indirect inversion, the seismic interpreters (the end users) would make qualitative statements that the seismic image does not look geologic.   These qualitative statements might be:

\begin{enumerate}
\item The geometry of the structure/stratigraphy does not look like what they expect based on tectonic/deposition.
\item Migration artifacts (cycle skipping or swings or static or multiples etc) cause geologic mis-interpretation.
\end{enumerate}

\subsection*{Some theoretical aspects}

In ''Seismic solvability problems'' (Lau and Yin, 2012), we showed that group theory (invertible system) is rare compared to semigroup theory (non-invertible or partially invertible system).  Group theory is ''nice'' and has symmetry.  Semigroup theory is ''messy'' but describes field seismic data realistically.  Information in general experiences ''loss'' and ''attenuation'' due to earth properties which are not invertible.  

One application of semigroup is diffusion semigroup.  It tries to maintain some flavor of group theory with an infinitesimal generator like the identity in a group.  But it allows the operator to ''diffuse'' as in ''Diffusion semigroups: A diffusion-map approach to nonlinear decomposition of seismic data without predetermined basis'' which respects data geometry (Lau et al., 2009).

\section*{Conclusions}

We have presented certain theorems in this paper.  Using only seismic data, it is difficult to find unique solution in seismic inversion with direct or indirect methods.  We are by no means discouraging further research into seismic inversion.  But we should be watchful that it could lead to non-unique and sometimes non-geologic results due to inherent limitations.

It is worthwhile to examine the data with good diagnostics (QC plots) for quality control when we apply quantitative methods like direct/indirect inversion.  Then we should incorporate other prior information like well logs, gravity, magnetic, interpretation, geometric measures, etc.

\section*{Appendix} 

We need to have practical solutions to real data problems. One compromise between direct and indirect inversion was suggested by Gasparotto and Lau (2000). Instead of using direct inversion to obtain primaries as the end result, we could use the direct inversion to "motivate" estimation of multiples. Since we use statistical methods like adaptive subtraction to remove multiples, the modeled multiples could be significantly simplified for subtraction. More recent example was published by Ramirez, et al in the two papers (2017, Part I and Part II) for internal multiple attenuation. 

Another comprise is to remove noise before direct inversion. Nita et al demonstrated spurious events occur after direct inversion when noise is present in synthetic data. It is not clear in general how to remove noise in real data. It could become subjective and requires expertise in understanding the underlying signal. Denoising does not have a systematic approach and is still ad hoc.

\bibliographystyle{seg}  
\bibliography{lau_yin_arxiv2017dindsi}

\end{document}